\documentclass[aps,apl,preprint,superscriptaddress]{revtex4-1}

\usepackage[version=3]{mhchem} 
\usepackage{graphicx}
\usepackage{hyperref}
\newcommand{\tc}{Wm$^{-1}$K$^{-1}$ }

\begin{document}


\title{Suppression of lattice thermal conductivity by mass-conserving cation mutation in multi-component semiconductors}


\author{Taizo Shibuya}
\affiliation{Department of Mechanical Engineering, Keio University, Yokohama 223-8522, Japan}
\author{Jonathan M. Skelton}
\author{Adam J. Jackson}
\affiliation{Department of Chemistry, University of Bath, Claverton Down, Bath BA2 7AY, United Kingdom}
\author{Kenji Yasuoka}
\affiliation{Department of Mechanical Engineering, Keio University, Yokohama 223-8522, Japan}
\author{Atsushi Togo}
\affiliation{Elements Strategy Initiative for Structural Materials, Kyoto University, Kyoto 606-8501, Japan}
\author{Isao Tanaka}
\affiliation{Elements Strategy Initiative for Structural Materials, Kyoto University, Kyoto 606-8501, Japan}
\affiliation{Department of Materials Science and Engineering, Kyoto University, Kyoto 606-8501, Japan}
\author{Aron Walsh}
\affiliation{Department of Chemistry, University of Bath, Claverton Down, Bath BA2 7AY, United Kingdom}
\affiliation{Global E$^3$ Institute and Department of Materials Science and Engineering, Yonsei University, Seoul 120-749, Korea}
\email[]{a.walsh@bath.ac.uk}


\date{\today}

\begin{abstract}
%
%
In semiconductors almost all heat is conducted by phonons (lattice vibrations), which is limited by their quasi-particle lifetimes. Phonon-phonon interactions represent scattering mechanisms that produce thermal resistance. In thermoelectric materials, this resistance due to anharmonicity should be maximised for optimal performance. We use a first-principles lattice-dynamics approach to explore the changes in lattice dynamics across an isostructural series where the average atomic mass is conserved: ZnS to \ce{CuGaS2} to \ce{Cu2ZnGeS4}. Our results demonstrate an enhancement of phonon interactions in the multernary materials, and confirm that lattice thermal conductivity can be controlled independently of the average mass and local coordination environments. 
\end{abstract}


\maketitle

Direct heat to electricity conversion in a thermoelectric device represents an important  process for energy generation and efficiency.\cite{Row1995}
The physical principles and applications of thermoelectric power conversion are well established;
however,
current commercial devices are based on heavy metal tellurides, which have issues associated with cost, toxicity and element availability. 
A new generation of materials are required to support growth of this technology, 
ideally with performance superior to existing compounds.\cite{Gaultois2013}

There has already been decades of research focused on discovering materials with enhanced performance.
A set of standard guidelines have emerged in the field and are widely adopted. 
These include: 
high atomic weight (which led to PbTe and \ce{Bi2Te3});\cite{Goldsmid2002}
loosely bound ``rattling" atoms (e.g. clathrates);\cite{Tse1988}; 
low dimensionality (e.g. nanorods and quantum dots);\cite{Hicks1993}
and phonon localization (e.g. superlattices and alloys).\cite{Wang2015w}
Since hundreds of materials have been systematically screened, tried and tested, 
new ideas are required to move this field forward.\cite{Yang2013a,Yan2014a,Seko2015}

A critical factor underpinning thermoelectric performance is thermal conductivity,
which represents a thermal loss mechanism that should be minimised for high-performance.
All crystals have phonons (extended lattice vibrations), but the thermal conductivity is linked to their lifetime,
which can vary by several orders of magnitude between materials.\cite{Spitzer1970}
Both electrons and phonons can play a role in thermal conductivity, but for dielectrics and intrinsic semiconductors the latter are usually dominant. 
An ideal thermoelectric material will offer ``phonon-glass electron-crystal" behaviour:\cite{Sales1997} a crystalline system where phonons have short lifetimes (low thermal conductivity), 
while electron carriers have long lifetimes (high electrical conductivity). 

In this study we attempt to answer a simple question: \textit{can the lattice thermal conductivity of tetrahedral 
semiconductors be tuned without changing the local structure or average mass of the compound?}
The topic was inspired by the early work of Pamplin\cite{Pamplin1964} and Goodman\cite{Goodman1958} on cation cross-substitution to produce multi-component semiconductors, which was later applied to screening materials for photovoltaics,\cite{Chen2009} 
topological conductivity\cite{Chen2011a} and spintronics\cite{Chen2009a}. 
We consider the mutation from ZnS $\rightarrow$ \ce{CuGaS2} $\rightarrow$  \ce{Cu2ZnGeS4} where the average atomic mass 
and crystal structures are conserved. 
A systematic analysis, from direct computation of the phonon energies and lifetimes, demonstrates 
a large decrease in phonon lifetimes on transition from the binary to ternary compounds, but no 
further enhancement is found in the quaternary system.
Analysis of the results suggests avenues for lowering the thermal conductivity of semiconducting crystals. 

\textbf{Lattice thermal conductivity from first-principles.}
Within the harmonic approximation, phonons propagate indefinitely and the lattice thermal conductivity is formally infinite.
Even in the absence of structural defects and chemical impurities,
the interaction \textit{between} phonon modes leads to scattering processes, of which anharmonic three-phonon
interactions are considered to be dominant.
As stated by Ziman in his seminal monograph: 
\textit{``Knowledge of the magnitude of the anharmonic terms which generate the [phonon-phonon scattering]
processes is scanty, and can only be obtained by roundabout arguments from other general macroscopic phenomena"}
\cite{Ziman1960}.
Fifty years later, the study of many-phonon processes remains a daunting task.

By solving the phonon Boltzmann equation within the relaxation-time approximation (RTA), the lattice thermal conductivity tensor ($\kappa$) can be expressed succinctly as a sum over phonons of band index $\lambda$ and wavevector $q$:
\begin{equation}
\kappa = \sum_{q \lambda} C_{V, q \lambda}  \nu_{q \lambda} \otimes \nu _{q \lambda} \tau_{q \lambda} 
\label{eqn1} 
\end{equation}
%
where $C_V$ is the isochoric modal heat capacity, $\nu$ is the group velocity, and 
$\nu _{q \lambda} \tau_{q \lambda} = \Lambda_{q \lambda}$, the 
 phonon mean free path.
The single-mode relaxation time $\tau$ can be deduced from the imaginary part of the phonon self-energy, computed within many-body perturbation theory.

At high temperatures, $C_V$ approaches a constant, so the two critical parameters for lattice thermal conductivity are $\nu$ and $\Lambda$. 
The velocity $\nu$ is determined simply by the phonon dispersion with respect to reciprocal-space wavevector ($\frac{\partial \omega }{\partial q}$). 
$\Lambda$ depends on $\nu$, but also on the relaxation time, which is closely related to the phonon-phonon interaction strength and the distribution of frequencies in the phonon density of states, the latter determining the number of possible energy-conserving scattering events.

It has only recently become possible to directly compute many-phonon interactions from first-principles.\cite{Broido2007,Skelton2014,Togo2015}
Even though a robust infrastructure now exists, there is a large computational cost to performing the simulations for complex crystals. 
By calculating the three-phonon interaction strength ($\phi_{\lambda\lambda'\lambda''}$), $\tau$ can be computed 
up to second-order within many-body perturbation theory.
We employ the \textsc{Phonopy} and \textsc{Phono3py} packages\cite{Togo2015,Togo2015a} using \textsc{VASP}\cite{Kresse1996c} as the force calculator and the PBEsol\cite{Perdew2008a} exchange-correlation treatment within Kohn-Sham density functional theory (DFT). 
PBEsol was chosen as it has been found to be an excellent compromise between accuracy and computational cost for lattice dynamics.\cite{Skelton2015a}
For each material we consider the harmonic phonon dispersion, thermal expansion within the quasi-harmonic approximation (QHA), and finally lattice thermal conductivity within the RTA. 
The technical set-up and procedure is provided in the computational details section. 
It should be noted that an alternative approach to capture anharmonic interactions is through molecular dynamics simulations where, for example, the Green-Kubo method can be used to probe phonon lifetimes and thermal conductivity.\cite{Qiu2012}

\begin{figure}[h!]
    \includegraphics[]{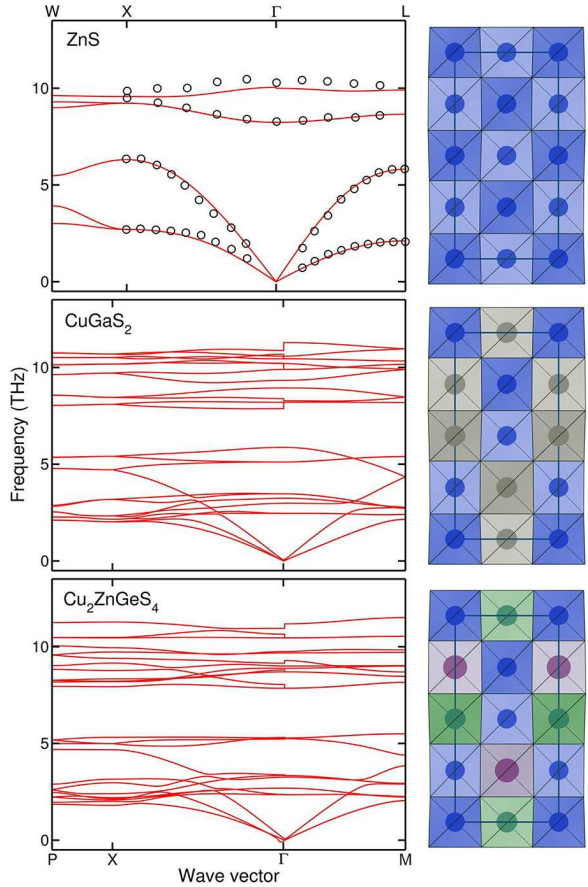}%
\caption{\label{f1} Phonon dispersion of ZnS,  \ce{CuGaS2} and \ce{Cu2ZnGeS4} as 
calculated within the harmonic approximation (PBEsol/DFT) at the equilibrium (athermal) 
lattice constants. LO-TO splitting of the $\Gamma$ point modes is included based on the calculated
dielectric tensors.
For ZnS, the dispersion determined from neutron scattering measurements\cite{Vagelatos1974} is overlaid. 
Illustrations of the crystal structures show the polyhedra coloured according to the cation sublattice. 
}
\end{figure}

\textbf{Zincblende ZnS.}
ZnS can crystallise in two polymorphs, but here we consider only zincblende (sphalerite), 
the most common face-centered-cubic form with space group type $F\bar{4}3m$ ($T_d$ symmetry). 
The primitive cell contains two lattice sites, and hence there are 6 phonon modes. 
At the zone-centre (\textit{q}=0), the $T_2$ (IR and Raman active) optic branch 
is split into lower-energy transverse (TO) and higher-energy longitudinal (LO) modes.

\begin{figure}[h]
    \includegraphics[]{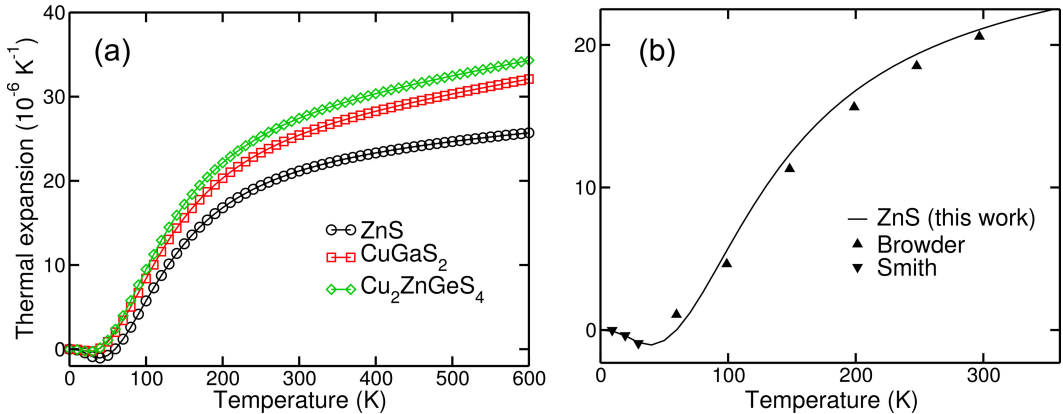}%
\caption{\label{f2}
(a) Temperature dependence of volumetric thermal expansion of ZnS, \ce{CuGaS2} and \ce{Cu2ZnGeS4} as 
calculated within the quasi-harmonic approximation (PBEsol/DFT). 
Note that similar negative thermal expansion at low temperatures is found in each material.
(b) Comparison between measurements (Browder\cite{Browder1977} and Smith\cite{Smith1975}) and simulation for ZnS.}
\end{figure}

In the first reported Raman measurement of ZnS crystals, 
the room temperature TO and LO modes
were measured at 276 and  351 cm$^{-1}$, respectively.\cite{Brafman1968}
The calculated values from the harmonic phonon dispersion (Figure \ref{f1})
are 275 and 335 cm$^{-1}$, in good agreement.
In addition, the phonon dispersion across 
the Brillouin zone agrees very well with neutron scattering measurements.\cite{Vagelatos1974}
%
The thermal expansion, calculated within the QHA, is plotted in Figure \ref{f2}.
The known negative thermal expansion at low temperatures is reproduced and the room temperature value of 
2 $\times 10^{-6}$ K is in good agreement with both experiment and reported lattice-dynamics calculations.\cite{Kagaya1987}

\textbf{From binary to ternary: \ce{CuGaS2}.}
The crystal structure of  \ce{CuGaS2} is a simple $1a\times1a\times2a$ 
supercell expansion of sphalerite with
the Cu and Ga atoms aligned along the (201) planes.
The chalcopyrite mineral structure with space group type $I\bar{4}2d$ ($D_{2d}$ symmetry) has eight atoms in the primitive cell 
and 24 phonon modes. 
Even at the Brillouin zone centre, the phonon structure is complicated with 
$A_1 + 2A_2 + 3B_1 + 4B_2 +7E$ vibrational branches, of which
only the $A_2$ modes are IR and Raman inactive.

The harmonic phonon dispersion (Figure \ref{f1}) 
is in good agreement with previous calculations\cite{Akdogan2002,Romero2011}.
The measured upper optic branch runs from 261 to 387 cm$^{-1}$ (collated in \cite{Akdogan2002}),
which matches well with our calculated values of 255 to 377 cm$^{-1}$.
%
%
The thermal expansion behaviour is found to be similar to ZnS, including the low-temperature lattice contraction. 
The linear expansion coefficient $\alpha$ = 25.4$\times 10 ^{-6}$K$^{-1}$ at 300 K is in agreement with the experimental measurements of  23 -- 27$\times 10 ^{-6}$K$^{-1}$\cite{Yamamoto1979,Bodnar1983}.

\textbf{From ternary to quaternary: \ce{Cu2ZnGeS4}.}
The kesterite crystal structure adopted by \ce{Cu2ZnGeS4} is closely related to chalcopyrite, 
with the addition of alternating Cu-Zn and Cu-Ge (001) planes.
The space group type is $I\bar{4}$ ($S_4$ symmetry) and there are again eight atoms in the primitive, giving rise to 24 phonon modes. 
The zone-centre irreducible representations are
$3A + 7B + 7E$. 
All modes Raman active, while only the B and E modes are IR active.

The phonon dispersion is shown in Figure \ref{f1}, and the optic branches run from 78 to 373 cm$^{-1}$.
The measured Raman modes run from 72 to 409 cm$^{-1}$, which were also found to be in good agreement with 
prior DFT calculations using the PBE functional.\cite{Guc2016} 
%


\begin{figure}[]
    \includegraphics[]{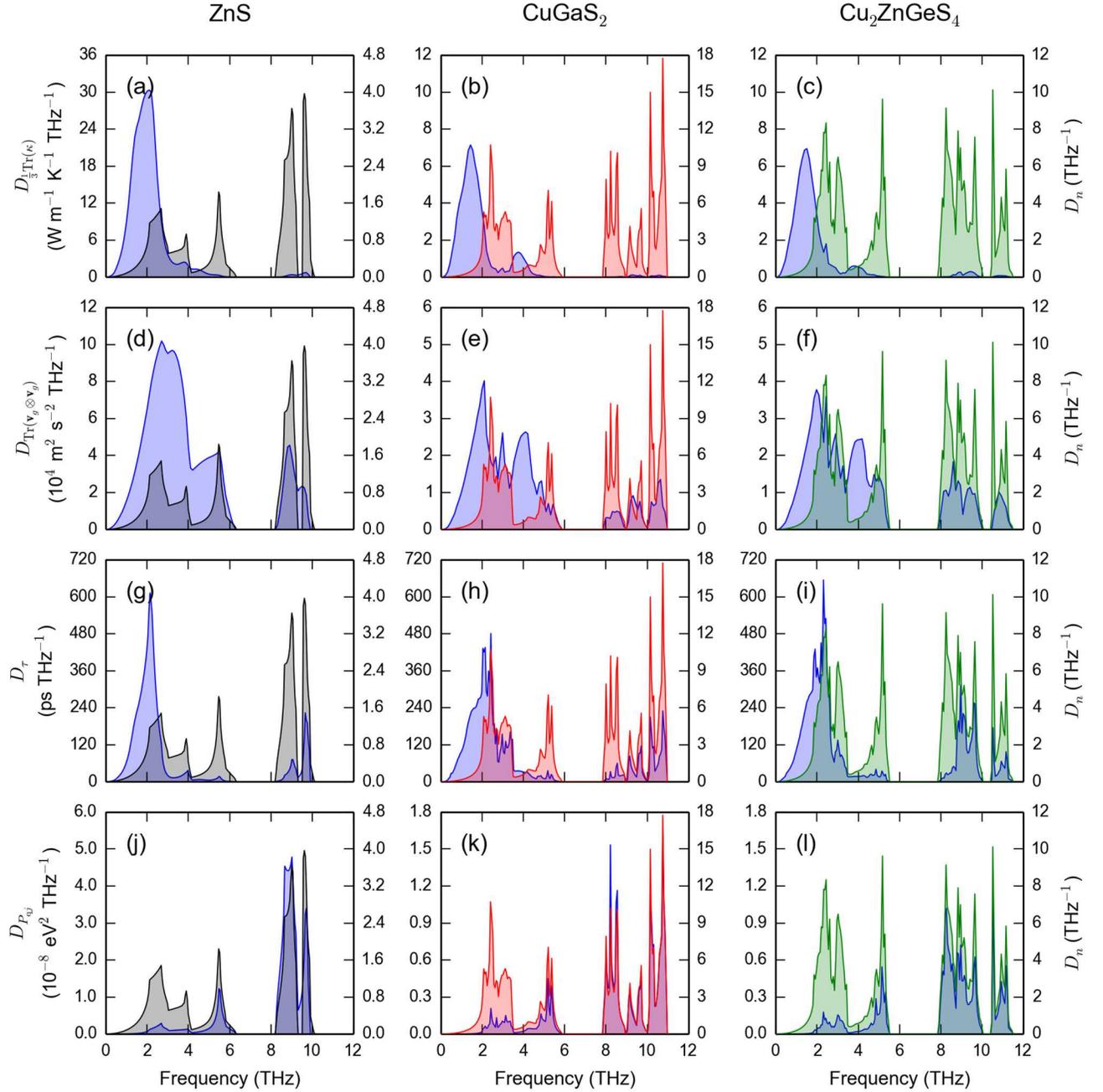}%
\caption{\label{f3} 
Density functions, $D$, showing the contribution of modes with different frequencies to the accumulated thermal conductivity in ZnS, \ce{CuGaS2} and \ce{Cu2ZnGeS4} (a-c). 
These are compared to similar functions computed from the trace of the group-velocity tensor products  (d-f), the phonon lifetimes  (g-i), and the averaged phonon-phonon interaction strengths  (j-l). 
On each plot, the functions are overlaid on the background phonon density of states (DoS) curves. A large value of  in a frequency band with a small DoS implies that those modes have a high value of the accumulated quantity (e.g. higher group velocity, longer lifetime).
}
\end{figure}

\textbf{Anharmonicity and thermal conductivity.}
To assess the modal contributions to the thermal transport, the room-temperature (300 K) modal thermal-conductivity tensors  
$\kappa_{\lambda}$ 
were integrated over the phonon Brillouin zone using the linear tetrahedron method. 
The frequency derivatives of the resulting accumulation functions were then computed, yielding a function analogous to a density of states (DoS) that quantifies the contribution of modes in different frequency bands to the overall transport. 
A similar procedure was followed for integrating the tensor products
$\nu _{q \lambda} \otimes \nu _{q \lambda}$, 
the phonon lifetimes
$\tau_{\lambda}$
, and the averaged phonon-phonon interaction strengths 
$P_{qj,\lambda}$ (see Ref. \cite{Togo2015} for further details).
With reference to the phonon DoS, these functions allow the modal contributions to the thermal conductivities to be interpreted in terms of mode lifetimes and group velocities (c.f. Eq. \ref{eqn1}).

The four sets of calculated functions are overlaid on the phonon DoS curves in Figure \ref{f3}. 
In all three materials, the majority of the heat transport is through modes up to $\sim$ 3 THz, which is due to a combination of long lifetimes and high group velocities. All three compounds also show a smaller secondary contribution to the thermal conductivity from modes between 3--4 THz with a comparable group velocity, but relatively shorter lifetimes. 
We note that the modal heat capacities
are saturated at 300 K, and the contribution of this term to  
$\kappa$
essentially mirrors the phonon DoS.

The data in Figure \ref{f3} shows that the reduction in thermal conductivity on going from ZnS to CGS and CZGS is attributable both to a reduction in the mode lifetimes, and a reduction in the group velocity, the latter of which can be attributed to the weaker bonding. For the most part, long-lived modes are associated with regions of the DoS in which 
the interaction strength
 is small. 
 It is worth noting that the interaction strength is considerably larger in ZnS than in CGS and CZGS, despite its longer phonon lifetimes and high thermal conductivity. This can be explained in terms of the conservation of energy: the more complex structures of CGS and CZGS lead to a broadening of the phonon DoS, particularly the high-frequency optic branches, which in turn provides more energy-conserving scattering channels that outweigh the weaker phonon-phonon interactions.

\begin{figure}[h]
    \includegraphics[]{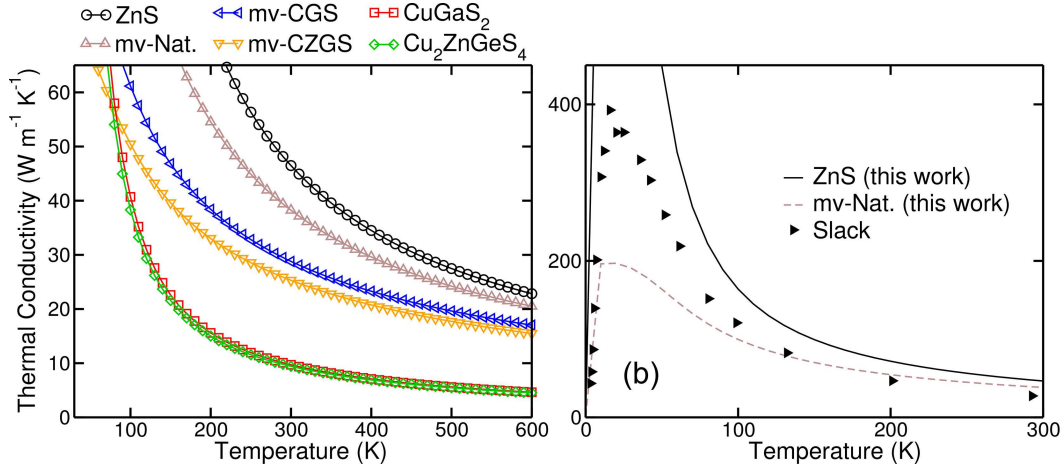}%
\caption{\label{f4}(a) Calculated lattice thermal conductivity of ZnS, \ce{CuGaS2} (CGS) and \ce{Cu2ZnGeS4} (CZGS) using harmonic force constants calculated within PBEsol/DFT. 
The isotropic average is shown for CGS and CZGS.
For ZnS, the effects of natural isotope variation (mv-Nat) and artificial isotope variation mirroring the ternary (mv-CGS) and quaternary (mv-CZGS) systems are also shown. 
(b) Comparison of the measured and calculated thermal conductivity of ZnS.
}
\end{figure}

Figure \ref{f4} shows the calculated thermal-conductivity curves for the three materials. The measured thermal conductivities for ZnS range from 360 \tc at 30 K to 27 \tc at 300 K,\cite{Madelung2003} the latter of which is within a factor of two of our calculated value. 
For \ce{CuGaS2} and \ce{Cu2ZnGeS4}, the isotropic average of the thermal conductivity 
$\kappa = \frac{1}{3}(\kappa_{xx} + \kappa_{yy} + \kappa_{zz})$
 is presented in Figure \ref{f4}. 
The anisotropy in the thermal conductivity of these tetragonal crystals is small, but non-negligible. 
A higher conductivity is found in the \textit{ab} plane ($\kappa_{xx}^{CGS}$ = $\kappa_{xx}^{CZGS}$ = 10.2 \tc at T = 300 K) than along the \textit{c} axis ($\kappa_{zz}^{CGS}$ = 8.8 \tc; $\kappa_{xx}^{CZGS}$ = 8.0 \tc at T = 300 K).

Although average mass is conserved for the three compounds considered here, the mass variance caused by the 2Zn $\rightarrow$ Cu + Ga mutation may result in additional scattering analogous to the anharmonic scattering found in disordered alloys.
We have tested this hypothesis by modifying ZnS with heavy and light isotopes in the same arrangement as the ternary and quaternary structures (labelled ``mv" in Figure \ref{f4}).
The 300 K thermal conductivity is reduced by 
18 \% for the natural isotope variation, 
38 \% for \ce{CuGaS2}-type variation,
and 46 \% for \ce{Cu2ZnGeS4}-type variation.
This mass variance contribution is significant, but a
 much weaker effect than the changes in the force constants caused by the chemical substitutions.  
In the actual calculations of \ce{CuGaS2} and  \ce{Cu2ZnGeS4}, the thermal conductivity is reduced by 78 \% in comparison 
to the value for ZnS at 300 K.

As noted above, acoustic phonon modes are responsible for conducting most of the heat, owing to their large group velocity and longer lifetime. While optical phonons (less disperse bands with lower $\nu$) do not directly contribute to thermal transport, they are important indirectly in determining mode lifetimes by their involvement in scattering processes. A clear change between the phonon DoS on 
 transition from the binary to multernary materials is that the optic branch of the DoS is widened. These modes mostly involve motion of the anion sub-lattice 
 --- animations of the modes are provided as supplementary information --- and the distribution of environments found when multiple cations are introduced causes a spread in frequency. The result is a higher probability of three-phonon interactions, limiting the lifetimes, and this, together with a reduction in the group velocity, serves to reduce the thermal conductivity.

In the 1970 overview by Spitzer\cite{Spitzer1970} on the thermal conductivity of semiconductors, an important observation was made:
\textit{``It is found that lattice thermal conductivity may be correlated rather reliably with crystal structure. In general, increasing coordination of the ions is associated with decreasing thermal conductivity".}
We have further shown that even for a fixed coordination number,  the composition can have a large impact.  
Now that the underlying contributions to thermal transport can be separated and quantified, 
deeper insights can be obtained into the structure-property relationships that can be used towards the rational engineering of thermal transport.

\textit{Computational details:}
For the density functional theory calculations, a kinetic energy cut-off of 450 eV for the plane wave-basis set was combined with a
reciprocal-space sampling equivalent to $8 \times 8 \times 8$ $k$-points for the zincblende primitive cell (i.e. $4 \times 4 \times 4$ for \ce{CuGaS2} and \ce{Cu2ZnGeS4}, respectively).
Projector augmented-wave (PAW) pseudopotentials\cite{Blochl1994} were employed.
A tolerance of 10$^{-8}$ eV was applied during the electronic minimization, and geometry optimisations were carried out to a force threshold of 10$^{-1}$ eV/\AA.
The precision of the charge-density grids was automatically chosen to avoid aliasing errors, and an additional support grid with $8 \times$ the number of points was used to evaluate the forces during the single-point force evaluations for the lattice-dynamics calculations.
The PAW projection was applied in reciprocal space (\textsc{LREAL = .FALSE.} in \textsc{VASP}).

For the lattice-dynamics calculations, second- and third-order force constants were computed in $2 \times 2 \times 2$ expansions of the primitive cells.
Additional calculations were carried out on cells with expansions and contractions of $\pm 3 \%$ about the the athermal equilibrium volume, in steps of 1 \%. 
The third-order calculations were performed using the 300 K volumes predicted from the quasi-harmonic approximation.

Phonon DoS curves and thermodynamic partition functions were computed using $\Gamma$-centred $q$-point grid with $48 \times 48 \times 48$ subdivisions to integrate the phonon Brillouin zones. 
The phonon lifetimes for ZnS and \ce{CuGaS2}/\ce{Cu2ZnGeS4} were calculated using $\Gamma$-centred $q$-point grids with $48 \times 48 \times 48$ and $16 \times 16 \times 16$ subdivisions, respectively, and the tetrahedron method was used for interpolation.

\clearpage

\textbf{Data Access Statement} 
The crystal structures and phonon data reported in this work are available in an online repository at \url{https://github.com/WMD-group/Phonons}, which can be processed using the \textsc{Phonopy} and \textsc{Phono3py} packages available from \url{http://atztogo.github.io/phonopy} and \url{http://atztogo.github.io/phono3py}.
The animations were made using \textsc{ascii-phonons}, available from \url{https://github.com/ajjackson/ascii-phonons}.

\begin{acknowledgments}
The calculations in this work used the facilities of the Supercomputer Center at thInse titute for Solid State Physics, University of Tokyo.
Some calculations were also performed on the UK Archer HPC facility, accessed through membership of the UK HPC Materials Chemistry Consortium, which is funded by EPSRC Grant No. EP/L000202.
We also made use of the Balena HPC facility at the University of Bath, which is maintained by Bath University Computing Services.
J.M.S. is funded by an EPSRC Programme Grant (grant no. EP/K004956/1). 
A.W. acknowledges support from the Royal Society and the ERC (grant no. 277757).
\end{acknowledgments}


\begin{thebibliography}{37}%
\makeatletter
\providecommand \@ifxundefined [1]{%
 \@ifx{#1\undefined}
}%
\providecommand \@ifnum [1]{%
 \ifnum #1\expandafter \@firstoftwo
 \else \expandafter \@secondoftwo
 \fi
}%
\providecommand \@ifx [1]{%
 \ifx #1\expandafter \@firstoftwo
 \else \expandafter \@secondoftwo
 \fi
}%
\providecommand \natexlab [1]{#1}%
\providecommand \enquote  [1]{``#1''}%
\providecommand \bibnamefont  [1]{#1}%
\providecommand \bibfnamefont [1]{#1}%
\providecommand \citenamefont [1]{#1}%
\providecommand \href@noop [0]{\@secondoftwo}%
\providecommand \href [0]{\begingroup \@sanitize@url \@href}%
\providecommand \@href[1]{\@@startlink{#1}\@@href}%
\providecommand \@@href[1]{\endgroup#1\@@endlink}%
\providecommand \@sanitize@url [0]{\catcode `\\12\catcode `\$12\catcode
  `\&12\catcode `\#12\catcode `\^12\catcode `\_12\catcode `\%12\relax}%
\providecommand \@@startlink[1]{}%
\providecommand \@@endlink[0]{}%
\providecommand \url  [0]{\begingroup\@sanitize@url \@url }%
\providecommand \@url [1]{\endgroup\@href {#1}{\urlprefix }}%
\providecommand \urlprefix  [0]{URL }%
\providecommand \Eprint [0]{\href }%
\providecommand \doibase [0]{http://dx.doi.org/}%
\providecommand \selectlanguage [0]{\@gobble}%
\providecommand \bibinfo  [0]{\@secondoftwo}%
\providecommand \bibfield  [0]{\@secondoftwo}%
\providecommand \translation [1]{[#1]}%
\providecommand \BibitemOpen [0]{}%
\providecommand \bibitemStop [0]{}%
\providecommand \bibitemNoStop [0]{.\EOS\space}%
\providecommand \EOS [0]{\spacefactor3000\relax}%
\providecommand \BibitemShut  [1]{\csname bibitem#1\endcsname}%
\let\auto@bib@innerbib\@empty
\bibitem [{\citenamefont {Rowe}(1995)}]{Row1995}%
  \BibitemOpen
  \bibfield  {author} {\bibinfo {author} {\bibfnamefont {D.~M.}\ \bibnamefont
  {Rowe}},\ }\href
  {https://books.google.com/books?hl=en{\&}lr={\&}id=Crtjc-luHlEC{\&}pgis=1}
  {\emph {\bibinfo {title} {{CRC Handbook of Thermoelectrics}}}}\ (\bibinfo
  {publisher} {CRC Press},\ \bibinfo {year} {1995})\ p.\ \bibinfo {pages}
  {701}\BibitemShut {NoStop}%
\bibitem [{\citenamefont {Gaultois}\ \emph {et~al.}(2013)\citenamefont
  {Gaultois}, \citenamefont {Sparks}, \citenamefont {Borg}, \citenamefont
  {Seshadri}, \citenamefont {Bonificio},\ and\ \citenamefont
  {Clarke}}]{Gaultois2013}%
  \BibitemOpen
  \bibfield  {author} {\bibinfo {author} {\bibfnamefont {M.~W.}\ \bibnamefont
  {Gaultois}}, \bibinfo {author} {\bibfnamefont {T.~D.}\ \bibnamefont
  {Sparks}}, \bibinfo {author} {\bibfnamefont {C.~K.~H.}\ \bibnamefont {Borg}},
  \bibinfo {author} {\bibfnamefont {R.}~\bibnamefont {Seshadri}}, \bibinfo
  {author} {\bibfnamefont {W.~D.}\ \bibnamefont {Bonificio}}, \ and\ \bibinfo
  {author} {\bibfnamefont {D.~R.}\ \bibnamefont {Clarke}},\ }\href {\doibase
  10.1021/cm400893e} {\bibfield  {journal} {\bibinfo  {journal} {Chem. Mater.}\
  }\textbf {\bibinfo {volume} {25}},\ \bibinfo {pages} {2911} (\bibinfo {year}
  {2013})}\BibitemShut {NoStop}%
\bibitem [{\citenamefont {Goldsmid}\ and\ \citenamefont
  {Douglas}(1954)}]{Goldsmid2002}%
  \BibitemOpen
  \bibfield  {author} {\bibinfo {author} {\bibfnamefont {H.~J.}\ \bibnamefont
  {Goldsmid}}\ and\ \bibinfo {author} {\bibfnamefont {R.~W.}\ \bibnamefont
  {Douglas}},\ }\href {\doibase 10.1088/0508-3443/5/12/513} {\bibfield
  {journal} {\bibinfo  {journal} {Br. J. Appl. Phys.}\ }\textbf {\bibinfo
  {volume} {5}},\ \bibinfo {pages} {458} (\bibinfo {year} {1954})}\BibitemShut
  {NoStop}%
\bibitem [{\citenamefont {Tse}\ and\ \citenamefont {White}(1988)}]{Tse1988}%
  \BibitemOpen
  \bibfield  {author} {\bibinfo {author} {\bibfnamefont {J.~S.}\ \bibnamefont
  {Tse}}\ and\ \bibinfo {author} {\bibfnamefont {M.~A.}\ \bibnamefont
  {White}},\ }\href {\doibase 10.1021/j100328a036} {\bibfield  {journal}
  {\bibinfo  {journal} {J. Phys. Chem.}\ }\textbf {\bibinfo {volume} {92}},\
  \bibinfo {pages} {5006} (\bibinfo {year} {1988})}\BibitemShut {NoStop}%
\bibitem [{\citenamefont {Hicks}\ and\ \citenamefont
  {Dresselhaus}(1993)}]{Hicks1993}%
  \BibitemOpen
  \bibfield  {author} {\bibinfo {author} {\bibfnamefont {L.~D.}\ \bibnamefont
  {Hicks}}\ and\ \bibinfo {author} {\bibfnamefont {M.~S.}\ \bibnamefont
  {Dresselhaus}},\ }\href {\doibase 10.1103/PhysRevB.47.16631} {\bibfield
  {journal} {\bibinfo  {journal} {Phys. Rev. B}\ }\textbf {\bibinfo {volume}
  {47}},\ \bibinfo {pages} {16631} (\bibinfo {year} {1993})}\BibitemShut
  {NoStop}%
\bibitem [{\citenamefont {Wang}\ \emph {et~al.}(2015)\citenamefont {Wang},
  \citenamefont {Gu},\ and\ \citenamefont {Ruan}}]{Wang2015w}%
  \BibitemOpen
  \bibfield  {author} {\bibinfo {author} {\bibfnamefont {Y.}~\bibnamefont
  {Wang}}, \bibinfo {author} {\bibfnamefont {C.}~\bibnamefont {Gu}}, \ and\
  \bibinfo {author} {\bibfnamefont {X.}~\bibnamefont {Ruan}},\ }\href {\doibase
  10.1063/1.4913319} {\bibfield  {journal} {\bibinfo  {journal} {Appl. Phys.
  Lett.}\ }\textbf {\bibinfo {volume} {106}},\ \bibinfo {pages} {073104}
  (\bibinfo {year} {2015})}\BibitemShut {NoStop}%
\bibitem [{\citenamefont {Yang}\ \emph {et~al.}(2013)\citenamefont {Yang},
  \citenamefont {Yip},\ and\ \citenamefont {Jen}}]{Yang2013a}%
  \BibitemOpen
  \bibfield  {author} {\bibinfo {author} {\bibfnamefont {J.}~\bibnamefont
  {Yang}}, \bibinfo {author} {\bibfnamefont {H.-L.}\ \bibnamefont {Yip}}, \
  and\ \bibinfo {author} {\bibfnamefont {A.~K.-Y.}\ \bibnamefont {Jen}},\
  }\href {\doibase 10.1002/aenm.201200514} {\bibfield  {journal} {\bibinfo
  {journal} {Adv. Energy Mater.}\ }\textbf {\bibinfo {volume} {3}},\ \bibinfo
  {pages} {549} (\bibinfo {year} {2013})}\BibitemShut {NoStop}%
\bibitem [{\citenamefont {Yan}\ \emph {et~al.}(2015)\citenamefont {Yan},
  \citenamefont {Gorai}, \citenamefont {Ortiz}, \citenamefont {Miller},
  \citenamefont {Barnett}, \citenamefont {Mason}, \citenamefont
  {Stevanovi{\'{c}}},\ and\ \citenamefont {Toberer}}]{Yan2014a}%
  \BibitemOpen
  \bibfield  {author} {\bibinfo {author} {\bibfnamefont {J.}~\bibnamefont
  {Yan}}, \bibinfo {author} {\bibfnamefont {P.}~\bibnamefont {Gorai}}, \bibinfo
  {author} {\bibfnamefont {B.}~\bibnamefont {Ortiz}}, \bibinfo {author}
  {\bibfnamefont {S.}~\bibnamefont {Miller}}, \bibinfo {author} {\bibfnamefont
  {S.~A.}\ \bibnamefont {Barnett}}, \bibinfo {author} {\bibfnamefont
  {T.}~\bibnamefont {Mason}}, \bibinfo {author} {\bibfnamefont
  {V.}~\bibnamefont {Stevanovi{\'{c}}}}, \ and\ \bibinfo {author}
  {\bibfnamefont {E.~S.}\ \bibnamefont {Toberer}},\ }\href {\doibase
  10.1039/C4EE03157A} {\bibfield  {journal} {\bibinfo  {journal} {Energy
  Environ. Sci.}\ }\textbf {\bibinfo {volume} {8}},\ \bibinfo {pages} {983}
  (\bibinfo {year} {2015})}\BibitemShut {NoStop}%
\bibitem [{\citenamefont {Seko}\ \emph {et~al.}(2015)\citenamefont {Seko},
  \citenamefont {Togo}, \citenamefont {Hayashi}, \citenamefont {Tsuda},
  \citenamefont {Chaput},\ and\ \citenamefont {Tanaka}}]{Seko2015}%
  \BibitemOpen
  \bibfield  {author} {\bibinfo {author} {\bibfnamefont {A.}~\bibnamefont
  {Seko}}, \bibinfo {author} {\bibfnamefont {A.}~\bibnamefont {Togo}}, \bibinfo
  {author} {\bibfnamefont {H.}~\bibnamefont {Hayashi}}, \bibinfo {author}
  {\bibfnamefont {K.}~\bibnamefont {Tsuda}}, \bibinfo {author} {\bibfnamefont
  {L.}~\bibnamefont {Chaput}}, \ and\ \bibinfo {author} {\bibfnamefont
  {I.}~\bibnamefont {Tanaka}},\ }\href {\doibase
  10.1103/PhysRevLett.115.205901} {\bibfield  {journal} {\bibinfo  {journal}
  {Phys. Rev. Lett.}\ }\textbf {\bibinfo {volume} {115}},\ \bibinfo {pages} {1}
  (\bibinfo {year} {2015})}\BibitemShut {NoStop}%
\bibitem [{\citenamefont {Spitzer}(1970)}]{Spitzer1970}%
  \BibitemOpen
  \bibfield  {author} {\bibinfo {author} {\bibfnamefont {D.~P.}\ \bibnamefont
  {Spitzer}},\ }\href {\doibase 10.1016/0022-3697(70)90284-2} {\bibfield
  {journal} {\bibinfo  {journal} {J. Phys. Chem. Solids}\ }\textbf {\bibinfo
  {volume} {31}},\ \bibinfo {pages} {19} (\bibinfo {year} {1970})}\BibitemShut
  {NoStop}%
\bibitem [{\citenamefont {Sales}\ \emph {et~al.}(1997)\citenamefont {Sales},
  \citenamefont {Mandrus}, \citenamefont {Chakoumakos}, \citenamefont
  {Keppens},\ and\ \citenamefont {Thompson}}]{Sales1997}%
  \BibitemOpen
  \bibfield  {author} {\bibinfo {author} {\bibfnamefont {B.~C.}\ \bibnamefont
  {Sales}}, \bibinfo {author} {\bibfnamefont {D.}~\bibnamefont {Mandrus}},
  \bibinfo {author} {\bibfnamefont {B.~C.}\ \bibnamefont {Chakoumakos}},
  \bibinfo {author} {\bibfnamefont {V.}~\bibnamefont {Keppens}}, \ and\
  \bibinfo {author} {\bibfnamefont {J.~R.}\ \bibnamefont {Thompson}},\ }\href
  {\doibase 10.1103/PhysRevB.56.15081} {\bibfield  {journal} {\bibinfo
  {journal} {Phys. Rev. B}\ }\textbf {\bibinfo {volume} {56}},\ \bibinfo
  {pages} {15081} (\bibinfo {year} {1997})}\BibitemShut {NoStop}%
\bibitem [{\citenamefont {Pamplin}(1964)}]{Pamplin1964}%
  \BibitemOpen
  \bibfield  {author} {\bibinfo {author} {\bibfnamefont {B.}~\bibnamefont
  {Pamplin}},\ }\href {\doibase 10.1016/0022-3697(64)90176-3} {\bibfield
  {journal} {\bibinfo  {journal} {J. Phys. Chem. Solids}\ }\textbf {\bibinfo
  {volume} {25}},\ \bibinfo {pages} {675} (\bibinfo {year} {1964})}\BibitemShut
  {NoStop}%
\bibitem [{\citenamefont {Goodman}(1958)}]{Goodman1958}%
  \BibitemOpen
  \bibfield  {author} {\bibinfo {author} {\bibfnamefont {C.~H.~L.}\
  \bibnamefont {Goodman}},\ }\href {\doibase 10.1016/0022-3697(58)90050-7}
  {\bibfield  {journal} {\bibinfo  {journal} {J. Phys. Chem. Solids}\ }\textbf
  {\bibinfo {volume} {6}},\ \bibinfo {pages} {305} (\bibinfo {year}
  {1958})}\BibitemShut {NoStop}%
\bibitem [{\citenamefont {Chen}\ \emph
  {et~al.}(2009{\natexlab{a}})\citenamefont {Chen}, \citenamefont {Gong},
  \citenamefont {Walsh},\ and\ \citenamefont {Wei}}]{Chen2009}%
  \BibitemOpen
  \bibfield  {author} {\bibinfo {author} {\bibfnamefont {S.}~\bibnamefont
  {Chen}}, \bibinfo {author} {\bibfnamefont {X.~G.}\ \bibnamefont {Gong}},
  \bibinfo {author} {\bibfnamefont {A.}~\bibnamefont {Walsh}}, \ and\ \bibinfo
  {author} {\bibfnamefont {S.-H.}\ \bibnamefont {Wei}},\ }\href {\doibase
  10.1103/PhysRevB.79.165211} {\bibfield  {journal} {\bibinfo  {journal} {Phys.
  Rev. B}\ }\textbf {\bibinfo {volume} {79}},\ \bibinfo {pages} {165211}
  (\bibinfo {year} {2009}{\natexlab{a}})}\BibitemShut {NoStop}%
\bibitem [{\citenamefont {Chen}\ \emph {et~al.}(2011)\citenamefont {Chen},
  \citenamefont {Gong}, \citenamefont {Duan}, \citenamefont {Zhu},
  \citenamefont {Chu}, \citenamefont {Walsh}, \citenamefont {Yao},
  \citenamefont {Ma},\ and\ \citenamefont {Wei}}]{Chen2011a}%
  \BibitemOpen
  \bibfield  {author} {\bibinfo {author} {\bibfnamefont {S.}~\bibnamefont
  {Chen}}, \bibinfo {author} {\bibfnamefont {X.~G.}\ \bibnamefont {Gong}},
  \bibinfo {author} {\bibfnamefont {C.-G.}\ \bibnamefont {Duan}}, \bibinfo
  {author} {\bibfnamefont {Z.-Q.}\ \bibnamefont {Zhu}}, \bibinfo {author}
  {\bibfnamefont {J.-H.}\ \bibnamefont {Chu}}, \bibinfo {author} {\bibfnamefont
  {A.}~\bibnamefont {Walsh}}, \bibinfo {author} {\bibfnamefont {Y.-G.}\
  \bibnamefont {Yao}}, \bibinfo {author} {\bibfnamefont {J.}~\bibnamefont
  {Ma}}, \ and\ \bibinfo {author} {\bibfnamefont {S.-H.}\ \bibnamefont {Wei}},\
  }\href {http://journals.aps.org/prb/abstract/10.1103/PhysRevB.83.245202}
  {\bibfield  {journal} {\bibinfo  {journal} {Phys. Rev. B}\ }\textbf {\bibinfo
  {volume} {83}},\ \bibinfo {pages} {245202} (\bibinfo {year}
  {2011})}\BibitemShut {NoStop}%
\bibitem [{\citenamefont {Chen}\ \emph
  {et~al.}(2009{\natexlab{b}})\citenamefont {Chen}, \citenamefont {Yin},
  \citenamefont {Yang}, \citenamefont {Gong}, \citenamefont {Walsh},\ and\
  \citenamefont {Wei}}]{Chen2009a}%
  \BibitemOpen
  \bibfield  {author} {\bibinfo {author} {\bibfnamefont {S.}~\bibnamefont
  {Chen}}, \bibinfo {author} {\bibfnamefont {W.-J.}\ \bibnamefont {Yin}},
  \bibinfo {author} {\bibfnamefont {J.-H.}\ \bibnamefont {Yang}}, \bibinfo
  {author} {\bibfnamefont {X.~G.}\ \bibnamefont {Gong}}, \bibinfo {author}
  {\bibfnamefont {A.}~\bibnamefont {Walsh}}, \ and\ \bibinfo {author}
  {\bibfnamefont {S.-H.}\ \bibnamefont {Wei}},\ }\href {\doibase
  10.1063/1.3193662} {\bibfield  {journal} {\bibinfo  {journal} {Appl. Phys.
  Lett.}\ }\textbf {\bibinfo {volume} {95}},\ \bibinfo {pages} {052102}
  (\bibinfo {year} {2009}{\natexlab{b}})}\BibitemShut {NoStop}%
\bibitem [{\citenamefont {Ziman}(1960)}]{Ziman1960}%
  \BibitemOpen
  \bibfield  {author} {\bibinfo {author} {\bibfnamefont {J.}~\bibnamefont
  {Ziman}},\ }\href
  {https://books.google.com/books?hl=en{\&}lr={\&}id=UtEy63pjngsC{\&}pgis=1}
  {\emph {\bibinfo {title} {{Electrons and Phonons: The Theory of Transport
  Phenomena in Solids}}}}\ (\bibinfo  {publisher} {OUP Oxford},\ \bibinfo
  {year} {1960})\ p.\ \bibinfo {pages} {554}\BibitemShut {NoStop}%
\bibitem [{\citenamefont {Broido}\ \emph {et~al.}(2007)\citenamefont {Broido},
  \citenamefont {Malorny}, \citenamefont {Birner}, \citenamefont {Mingo},\ and\
  \citenamefont {Stewart}}]{Broido2007}%
  \BibitemOpen
  \bibfield  {author} {\bibinfo {author} {\bibfnamefont {D.~A.}\ \bibnamefont
  {Broido}}, \bibinfo {author} {\bibfnamefont {M.}~\bibnamefont {Malorny}},
  \bibinfo {author} {\bibfnamefont {G.}~\bibnamefont {Birner}}, \bibinfo
  {author} {\bibfnamefont {N.}~\bibnamefont {Mingo}}, \ and\ \bibinfo {author}
  {\bibfnamefont {D.~A.}\ \bibnamefont {Stewart}},\ }\href {\doibase
  10.1063/1.2822891} {\bibfield  {journal} {\bibinfo  {journal} {Appl. Phys.
  Lett.}\ }\textbf {\bibinfo {volume} {91}},\ \bibinfo {pages} {231922}
  (\bibinfo {year} {2007})}\BibitemShut {NoStop}%
\bibitem [{\citenamefont {Skelton}\ \emph {et~al.}(2014)\citenamefont
  {Skelton}, \citenamefont {Parker}, \citenamefont {Togo}, \citenamefont
  {Tanaka},\ and\ \citenamefont {Walsh}}]{Skelton2014}%
  \BibitemOpen
  \bibfield  {author} {\bibinfo {author} {\bibfnamefont {J.~M.}\ \bibnamefont
  {Skelton}}, \bibinfo {author} {\bibfnamefont {S.~C.}\ \bibnamefont {Parker}},
  \bibinfo {author} {\bibfnamefont {A.}~\bibnamefont {Togo}}, \bibinfo {author}
  {\bibfnamefont {I.}~\bibnamefont {Tanaka}}, \ and\ \bibinfo {author}
  {\bibfnamefont {A.}~\bibnamefont {Walsh}},\ }\href {\doibase
  10.1103/PhysRevB.89.205203} {\bibfield  {journal} {\bibinfo  {journal} {Phys.
  Rev. B}\ }\textbf {\bibinfo {volume} {89}},\ \bibinfo {pages} {205203}
  (\bibinfo {year} {2014})}\BibitemShut {NoStop}%
\bibitem [{\citenamefont {Togo}\ \emph {et~al.}(2015)\citenamefont {Togo},
  \citenamefont {Chaput},\ and\ \citenamefont {Tanaka}}]{Togo2015}%
  \BibitemOpen
  \bibfield  {author} {\bibinfo {author} {\bibfnamefont {A.}~\bibnamefont
  {Togo}}, \bibinfo {author} {\bibfnamefont {L.}~\bibnamefont {Chaput}}, \ and\
  \bibinfo {author} {\bibfnamefont {I.}~\bibnamefont {Tanaka}},\ }\href
  {\doibase 10.1103/PhysRevB.91.094306} {\bibfield  {journal} {\bibinfo
  {journal} {Phys. Rev. B}\ }\textbf {\bibinfo {volume} {91}},\ \bibinfo
  {pages} {094306} (\bibinfo {year} {2015})}\BibitemShut {NoStop}%
\bibitem [{\citenamefont {Togo}\ and\ \citenamefont
  {Tanaka}(2015)}]{Togo2015a}%
  \BibitemOpen
  \bibfield  {author} {\bibinfo {author} {\bibfnamefont {A.}~\bibnamefont
  {Togo}}\ and\ \bibinfo {author} {\bibfnamefont {I.}~\bibnamefont {Tanaka}},\
  }\href {\doibase 10.1016/j.scriptamat.2015.07.021} {\bibfield  {journal}
  {\bibinfo  {journal} {Scr. Mater.}\ }\textbf {\bibinfo {volume} {108}},\
  \bibinfo {pages} {1} (\bibinfo {year} {2015})}\BibitemShut {NoStop}%
\bibitem [{\citenamefont {Kresse}\ and\ \citenamefont
  {Furthm{\"{u}}ller}(1996)}]{Kresse1996c}%
  \BibitemOpen
  \bibfield  {author} {\bibinfo {author} {\bibfnamefont {G.}~\bibnamefont
  {Kresse}}\ and\ \bibinfo {author} {\bibfnamefont {J.}~\bibnamefont
  {Furthm{\"{u}}ller}},\ }\href {\doibase 10.1103/PhysRevB.54.11169} {\bibfield
   {journal} {\bibinfo  {journal} {Phys. Rev. B}\ }\textbf {\bibinfo {volume}
  {54}},\ \bibinfo {pages} {11169} (\bibinfo {year} {1996})}\BibitemShut
  {NoStop}%
\bibitem [{\citenamefont {Perdew}\ \emph {et~al.}(2008)\citenamefont {Perdew},
  \citenamefont {Ruzsinszky}, \citenamefont {Csonka}, \citenamefont {Vydrov},
  \citenamefont {Scuseria}, \citenamefont {Constantin}, \citenamefont {Zhou},\
  and\ \citenamefont {Burke}}]{Perdew2008a}%
  \BibitemOpen
  \bibfield  {author} {\bibinfo {author} {\bibfnamefont {J.~P.}\ \bibnamefont
  {Perdew}}, \bibinfo {author} {\bibfnamefont {A.}~\bibnamefont {Ruzsinszky}},
  \bibinfo {author} {\bibfnamefont {G.~I.}\ \bibnamefont {Csonka}}, \bibinfo
  {author} {\bibfnamefont {O.~A.}\ \bibnamefont {Vydrov}}, \bibinfo {author}
  {\bibfnamefont {G.~E.}\ \bibnamefont {Scuseria}}, \bibinfo {author}
  {\bibfnamefont {L.~A.}\ \bibnamefont {Constantin}}, \bibinfo {author}
  {\bibfnamefont {X.}~\bibnamefont {Zhou}}, \ and\ \bibinfo {author}
  {\bibfnamefont {K.}~\bibnamefont {Burke}},\ }\href {\doibase
  10.1103/PhysRevLett.100.136406} {\bibfield  {journal} {\bibinfo  {journal}
  {Phys. Rev. Lett.}\ }\textbf {\bibinfo {volume} {100}},\ \bibinfo {pages}
  {136406} (\bibinfo {year} {2008})}\BibitemShut {NoStop}%
\bibitem [{\citenamefont {Skelton}\ \emph {et~al.}(2015)\citenamefont
  {Skelton}, \citenamefont {Tiana}, \citenamefont {Parker}, \citenamefont
  {Togo}, \citenamefont {Tanaka},\ and\ \citenamefont {Walsh}}]{Skelton2015a}%
  \BibitemOpen
  \bibfield  {author} {\bibinfo {author} {\bibfnamefont {J.~M.}\ \bibnamefont
  {Skelton}}, \bibinfo {author} {\bibfnamefont {D.}~\bibnamefont {Tiana}},
  \bibinfo {author} {\bibfnamefont {S.~C.}\ \bibnamefont {Parker}}, \bibinfo
  {author} {\bibfnamefont {A.}~\bibnamefont {Togo}}, \bibinfo {author}
  {\bibfnamefont {I.}~\bibnamefont {Tanaka}}, \ and\ \bibinfo {author}
  {\bibfnamefont {A.}~\bibnamefont {Walsh}},\ }\href {\doibase
  10.1063/1.4928058} {\bibfield  {journal} {\bibinfo  {journal} {J. Chem.
  Phys.}\ }\textbf {\bibinfo {volume} {143}},\ \bibinfo {pages} {064710}
  (\bibinfo {year} {2015})}\BibitemShut {NoStop}%
\bibitem [{\citenamefont {Qiu}\ \emph {et~al.}(2012)\citenamefont {Qiu},
  \citenamefont {Bao}, \citenamefont {Zhang}, \citenamefont {Wu},\ and\
  \citenamefont {Ruan}}]{Qiu2012}%
  \BibitemOpen
  \bibfield  {author} {\bibinfo {author} {\bibfnamefont {B.}~\bibnamefont
  {Qiu}}, \bibinfo {author} {\bibfnamefont {H.}~\bibnamefont {Bao}}, \bibinfo
  {author} {\bibfnamefont {G.}~\bibnamefont {Zhang}}, \bibinfo {author}
  {\bibfnamefont {Y.}~\bibnamefont {Wu}}, \ and\ \bibinfo {author}
  {\bibfnamefont {X.}~\bibnamefont {Ruan}},\ }\href {\doibase
  10.1016/j.commatsci.2011.08.016} {\bibfield  {journal} {\bibinfo  {journal}
  {Comput. Mater. Sci.}\ }\textbf {\bibinfo {volume} {53}},\ \bibinfo {pages}
  {278} (\bibinfo {year} {2012})}\BibitemShut {NoStop}%
\bibitem [{\citenamefont {Vagelatos}\ \emph {et~al.}(1974)\citenamefont
  {Vagelatos}, \citenamefont {Wehe},\ and\ \citenamefont
  {King}}]{Vagelatos1974}%
  \BibitemOpen
  \bibfield  {author} {\bibinfo {author} {\bibfnamefont {N.}~\bibnamefont
  {Vagelatos}}, \bibinfo {author} {\bibfnamefont {D.}~\bibnamefont {Wehe}}, \
  and\ \bibinfo {author} {\bibfnamefont {J.~S.}\ \bibnamefont {King}},\
  }\href@noop {} {\bibfield  {journal} {\bibinfo  {journal} {J. Chem. Phys.}\
  }\textbf {\bibinfo {volume} {60}},\ \bibinfo {pages} {3613} (\bibinfo {year}
  {1974})}\BibitemShut {NoStop}%
\bibitem [{\citenamefont {Brafman}\ and\ \citenamefont
  {Mitra}(1968)}]{Brafman1968}%
  \BibitemOpen
  \bibfield  {author} {\bibinfo {author} {\bibfnamefont {O.}~\bibnamefont
  {Brafman}}\ and\ \bibinfo {author} {\bibfnamefont {S.~S.}\ \bibnamefont
  {Mitra}},\ }\href {\doibase 10.1103/PhysRev.171.931} {\bibfield  {journal}
  {\bibinfo  {journal} {Phys. Rev.}\ }\textbf {\bibinfo {volume} {171}},\
  \bibinfo {pages} {931} (\bibinfo {year} {1968})}\BibitemShut {NoStop}%
\bibitem [{\citenamefont {Kagaya}\ and\ \citenamefont
  {Soma}(1987)}]{Kagaya1987}%
  \BibitemOpen
  \bibfield  {author} {\bibinfo {author} {\bibfnamefont {H.-M.}\ \bibnamefont
  {Kagaya}}\ and\ \bibinfo {author} {\bibfnamefont {T.}~\bibnamefont {Soma}},\
  }\href {\doibase 10.1002/pssb.2221420210} {\bibfield  {journal} {\bibinfo
  {journal} {Phys. Status Solidi}\ }\textbf {\bibinfo {volume} {142}},\
  \bibinfo {pages} {411} (\bibinfo {year} {1987})}\BibitemShut {NoStop}%
\bibitem [{\citenamefont {Browder}\ and\ \citenamefont
  {Ballard}(1977)}]{Browder1977}%
  \BibitemOpen
  \bibfield  {author} {\bibinfo {author} {\bibfnamefont {J.~S.}\ \bibnamefont
  {Browder}}\ and\ \bibinfo {author} {\bibfnamefont {S.~S.}\ \bibnamefont
  {Ballard}},\ }\href {\doibase 10.1364/AO.16.003214} {\bibfield  {journal}
  {\bibinfo  {journal} {Appl. Opt.}\ }\textbf {\bibinfo {volume} {16}},\
  \bibinfo {pages} {3214} (\bibinfo {year} {1977})}\BibitemShut {NoStop}%
\bibitem [{\citenamefont {Smith}\ and\ \citenamefont
  {White}(1975)}]{Smith1975}%
  \BibitemOpen
  \bibfield  {author} {\bibinfo {author} {\bibfnamefont {T.~F.}\ \bibnamefont
  {Smith}}\ and\ \bibinfo {author} {\bibfnamefont {G.~K.}\ \bibnamefont
  {White}},\ }\href {\doibase 10.1088/0022-3719/8/13/012} {\bibfield  {journal}
  {\bibinfo  {journal} {J. Phys. C Solid State Phys.}\ }\textbf {\bibinfo
  {volume} {8}},\ \bibinfo {pages} {2031} (\bibinfo {year} {1975})}\BibitemShut
  {NoStop}%
\bibitem [{\citenamefont {Akdogan}\ and\ \citenamefont
  {Eryigit}(2002)}]{Akdogan2002}%
  \BibitemOpen
  \bibfield  {author} {\bibinfo {author} {\bibfnamefont {M.}~\bibnamefont
  {Akdogan}}\ and\ \bibinfo {author} {\bibfnamefont {R.}~\bibnamefont
  {Eryigit}},\ }\href {\doibase 10.1088/0953-8984/14/32/309} {\bibfield
  {journal} {\bibinfo  {journal} {J. Phys. Condens. Matter}\ }\textbf {\bibinfo
  {volume} {14}},\ \bibinfo {pages} {7493} (\bibinfo {year}
  {2002})}\BibitemShut {NoStop}%
\bibitem [{\citenamefont {Romero}\ \emph {et~al.}(2011)\citenamefont {Romero},
  \citenamefont {Cardona}, \citenamefont {Kremer}, \citenamefont {Lauck},
  \citenamefont {Siegle}, \citenamefont {Hoch}, \citenamefont {Munoz},\ and\
  \citenamefont {Schindler}}]{Romero2011}%
  \BibitemOpen
  \bibfield  {author} {\bibinfo {author} {\bibfnamefont {A.~H.}\ \bibnamefont
  {Romero}}, \bibinfo {author} {\bibfnamefont {M.}~\bibnamefont {Cardona}},
  \bibinfo {author} {\bibfnamefont {R.~K.}\ \bibnamefont {Kremer}}, \bibinfo
  {author} {\bibfnamefont {R.}~\bibnamefont {Lauck}}, \bibinfo {author}
  {\bibfnamefont {G.}~\bibnamefont {Siegle}}, \bibinfo {author} {\bibfnamefont
  {C.}~\bibnamefont {Hoch}}, \bibinfo {author} {\bibfnamefont {A.}~\bibnamefont
  {Munoz}}, \ and\ \bibinfo {author} {\bibfnamefont {A.}~\bibnamefont
  {Schindler}},\ }\href@noop {} {\bibfield  {journal} {\bibinfo  {journal}
  {Phys. Rev. B}\ }\textbf {\bibinfo {volume} {83}},\ \bibinfo {pages} {195208}
  (\bibinfo {year} {2011})}\BibitemShut {NoStop}%
\bibitem [{\citenamefont {Yamamoto}\ \emph {et~al.}(1979)\citenamefont
  {Yamamoto}, \citenamefont {Horinaka},\ and\ \citenamefont
  {Miyauchi}}]{Yamamoto1979}%
  \BibitemOpen
  \bibfield  {author} {\bibinfo {author} {\bibfnamefont {N.}~\bibnamefont
  {Yamamoto}}, \bibinfo {author} {\bibfnamefont {H.}~\bibnamefont {Horinaka}},
  \ and\ \bibinfo {author} {\bibfnamefont {T.}~\bibnamefont {Miyauchi}},\
  }\href@noop {} {\bibfield  {journal} {\bibinfo  {journal} {Jpn. J. Appl.
  Phys.}\ }\textbf {\bibinfo {volume} {18}},\ \bibinfo {pages} {255} (\bibinfo
  {year} {1979})}\BibitemShut {NoStop}%
\bibitem [{\citenamefont {Bodnar}\ and\ \citenamefont
  {Orlova}(1983)}]{Bodnar1983}%
  \BibitemOpen
  \bibfield  {author} {\bibinfo {author} {\bibfnamefont {I.~V.}\ \bibnamefont
  {Bodnar}}\ and\ \bibinfo {author} {\bibfnamefont {N.~S.}\ \bibnamefont
  {Orlova}},\ }\href@noop {} {\bibfield  {journal} {\bibinfo  {journal} {Phys.
  Status Solidi}\ }\textbf {\bibinfo {volume} {78}},\ \bibinfo {pages} {K59}
  (\bibinfo {year} {1983})}\BibitemShut {NoStop}%
\bibitem [{\citenamefont {Guc}\ \emph {et~al.}(2016)\citenamefont {Guc},
  \citenamefont {Litvinchuk}, \citenamefont {Levcenko}, \citenamefont {Valakh},
  \citenamefont {Bodnar}, \citenamefont {Dzhagan}, \citenamefont
  {Izquierdo-Roca}, \citenamefont {Arushanov},\ and\ \citenamefont
  {P{\'{e}}rez-Rodr{\'{i}}guez}}]{Guc2016}%
  \BibitemOpen
  \bibfield  {author} {\bibinfo {author} {\bibfnamefont {M.}~\bibnamefont
  {Guc}}, \bibinfo {author} {\bibfnamefont {A.~P.}\ \bibnamefont {Litvinchuk}},
  \bibinfo {author} {\bibfnamefont {S.}~\bibnamefont {Levcenko}}, \bibinfo
  {author} {\bibfnamefont {M.~Y.}\ \bibnamefont {Valakh}}, \bibinfo {author}
  {\bibfnamefont {I.~V.}\ \bibnamefont {Bodnar}}, \bibinfo {author}
  {\bibfnamefont {V.~M.}\ \bibnamefont {Dzhagan}}, \bibinfo {author}
  {\bibfnamefont {V.}~\bibnamefont {Izquierdo-Roca}}, \bibinfo {author}
  {\bibfnamefont {E.}~\bibnamefont {Arushanov}}, \ and\ \bibinfo {author}
  {\bibfnamefont {A.}~\bibnamefont {P{\'{e}}rez-Rodr{\'{i}}guez}},\ }\href
  {\doibase 10.1039/C5RA26844C} {\bibfield  {journal} {\bibinfo  {journal} {RSC
  Adv.}\ }\textbf {\bibinfo {volume} {6}},\ \bibinfo {pages} {13278} (\bibinfo
  {year} {2016})}\BibitemShut {NoStop}%
\bibitem [{\citenamefont {Madelung}(2003)}]{Madelung2003}%
  \BibitemOpen
  \bibfield  {author} {\bibinfo {author} {\bibfnamefont {O.~M.}\ \bibnamefont
  {Madelung}},\ }\href@noop {} {\emph {\bibinfo {title} {{Semiconductors: Data
  Handbook}}}},\ \bibinfo {edition} {3rd}\ ed.\ (\bibinfo  {publisher}
  {Springer},\ \bibinfo {address} {Berlin},\ \bibinfo {year} {2003})\ p.\
  \bibinfo {pages} {691}\BibitemShut {NoStop}%
\bibitem [{\citenamefont {Bl{\"{o}}chl}(1994)}]{Blochl1994}%
  \BibitemOpen
  \bibfield  {author} {\bibinfo {author} {\bibfnamefont {P.~E.}\ \bibnamefont
  {Bl{\"{o}}chl}},\ }\href {\doibase 10.1103/PhysRevB.50.17953} {\bibfield
  {journal} {\bibinfo  {journal} {Phys. Rev. B}\ }\textbf {\bibinfo {volume}
  {50}},\ \bibinfo {pages} {17953} (\bibinfo {year} {1994})}\BibitemShut
  {NoStop}%
\end{thebibliography}

%

\end{document}